\documentclass[a4paper]{article}%{ieeeaccess}
\usepackage{cite}
\usepackage{algorithm}
\usepackage{amsmath,amssymb,amsfonts}
\usepackage{algorithmic}
\usepackage[dvipdfmx]{graphicx}
\usepackage{color}
\usepackage{textcomp}
\usepackage{amsbsy}
  \usepackage{amsmath, amsfonts, mathrsfs, amsthm,here,colortbl, physics, ascmac, url}
\usepackage{latexsym}
\usepackage{amssymb}
\usepackage{amsbsy}
 \usepackage{comment}
\usepackage{subcaption}

\newtheorem{remark}{Remark}

\newcommand{\sgn}{\mathop{\mathrm{sgn}}\nolimits}

\definecolor{accessblue}{cmyk}{1, 0.3, 0, 0.2}
\definecolor{greycolor}{cmyk}{0,0,0,.8}
\def\BibTeX{{\rm B\kern-.05em{\sc i\kern-.025em b}\kern-.08em
T\kern-.1667em\lower.7ex\hbox{E}\kern-.125emX}}
\begin{document}
%\history{Date of publication xxxx 00, 0000, date of current version xxxx 00, 0000.}
%\doi{10.1109/ACCESS.2017.DOI}

\title{Robust Path Following Control for Vehicles with Uncertain Steering Torque Using Model Error Compensation}
\author{
Rentaro Iwai, Natsuki Hikasa and Hiroshi Okajima
}

%\address[1]{Graduate School of Science and Technology, Kumamoto University, Japan}
%\address[2]{Faculty of Environmental Engineering, the University of Kitakyushu, Japan}
%\address[3]{Osaka Institute of Technology, Japan}

%\address[2]{Graduate School of Engineering, Kyoto University, Kyoto, Japan}

%\tfootnote{This paragraph of the first footnote will contain support
%information, including sponsor and financial support acknowledgment. For
%example, ``This work was supported in part by the U.S. Department of
%Commerce under Grant BS123456.''}

%\markboth
%{Author \headeretal: Preparation of Papers for IEEE TRANSACTIONS and JOURNALS}
%{Author \headeretal: Preparation of Papers for IEEE TRANSACTIONS and JOURNALS}

%\corresp{Corresponding Author: H. Okajima (e-mail: okajima@cs.kumamoto-u.ac.jp)}

%\begin{abstract}
%This paper addresses a system identification for linear periodically time-varying plants in the discrete-time setting. A system identification algorithm for linear, periodically time-varying plants is introduced based on a cyclic reformulation and a state coordinate transformation of the cycled system. By using our system identification algorithm, the high-accuracy model of the periodically time-varying plant can be obtained without using specific periodic input signals. The effectiveness of the proposed algorithm is demonstrated with numerical examples. %linear periodically time-varying systems are useful. 
%\end{abstract}

%\begin{keywords}
%Cyclic reformulation, Time-varying systems, Subspace identification, and System identification.
%\end{keywords}

%\titlepgskip=-15pt

\begin{abstract}
This paper presents a robust path following control method for vehicles that explicitly incorporates steering torque dynamics into the control model. Unlike conventional approaches that treat the steering angle as a direct control input, this study models the steering angle as a state variable driven by a torque-proportional steering command against a speed- and angle-dependent resistance term. Since the resistance coefficient depends on road surface properties and is difficult to determine precisely, it is treated as an uncertain parameter. To compensate for the resulting model error, a Model Error Compensator (MEC) is employed as an add-on compensator that feeds back the discrepancy between the actual plant and a nominal model running in parallel, without requiring an explicit plant inverse or a specific canonical form. The zero dynamics arising from the path following formulation are formally analyzed, and it is shown that they are stable for any positive rear cornering power. Numerical simulations under systematic parameter mismatch conditions ($C/C_{M}=0.5$ to $2.0$) demonstrate that the proposed method reduces the maximum following error by more than 90\% compared to the conventional method without MEC throughout the tested mismatch range, and maintains practical tracking performance within a mismatch range of $0.75 \leq C/C_{M} \leq 1.25$. These results confirm that MEC effectively suppresses the influence of steering torque uncertainty, significantly enhancing path following robustness.
\end{abstract}

\maketitle

\section{INTRODUCTION}\label{section1}
In recent years, accidents caused by truck drivers' fatigue and the decline in physical and cognitive functions of elderly drivers have become significant societal concerns. Conventional studies indicate that over 90\% of traffic accidents are attributed to driver errors\cite{A1}. In response to these challenges, autonomous driving technology and advanced driver assistance systems (ADAS) have garnered considerable attention.

Autonomous driving offers several advantages, including the elimination of delays in perception, decision-making, and operation, as well as the reduction of human errors. Moreover, it enables highly precise control that surpasses human driving capabilities. Given these advantages, autonomous driving is expected to alleviate driver workload and enhance road safety. Consequently, extensive research and technological development are being conducted to facilitate its practical implementation with objectives such as accident prevention, traffic congestion mitigation, and improved passenger comfort and convenience.

To achieve autonomous vehicle operation, it is essential to develop control algorithms that autonomously execute the functions traditionally performed by human drivers, including steering, acceleration, and braking\cite{A2}. Among these, steering control corresponds to lateral vehicle operation, while speed control manages acceleration and braking. A fundamental requirement for autonomous vehicles is the ability to accurately follow a prescribed reference path, necessitating the design of a path following control algorithm.

Control methods for achieving path tracking can be broadly classified into trajectory tracking control and path following control. Trajectory tracking control ensures that the vehicle follows a time-dependent reference trajectory, meaning that the vehicle must reach a specific position at a given time \cite{A3,A4}. In contrast, path following control aims to align the vehicle's trajectory with a predefined reference path, without explicit time constraints. Since the objective of this study is to achieve precise alignment of the vehicle's trajectory with the reference path, we adopt a path following control approach.

Path following control has been widely studied for various control targets, with particular emphasis on its application to wheeled vehicles. Accurate path following is crucial not only for conventional automotive applications, such as autonomous driving and lane-keeping\cite{A5}, but also for a wide range of mobile systems, including agricultural machinery, construction vehicles, mobile robots, and automated guided vehicles (AGVs). In these systems, precise path following directly affects operational efficiency, safety, and task quality, making high-precision control essential. Path following formulations themselves also continue to evolve; for example, a projection-point-based formulation for path following along three-dimensional paths has recently been proposed \cite{A24}. Such three-dimensional formulations are also relevant to ground vehicles, since extending path following to uneven terrain naturally leads to three-dimensional target paths. However, vehicle motion control presents inherent challenges due to the kinematic constraints of conventional road vehicles, which limit lateral motion and impose nonholonomic restrictions \cite{A6}. Moreover, external disturbances such as variations in road friction and changes in vehicle load can influence the steering characteristics and make precise path following more challenging.

To address these challenges, various control approaches have been proposed that account for vehicle dynamics while ensuring accurate path following. Conventional studies include methods based on simplified vehicle models that neglect tire slip angles, which are computationally efficient and easier to implement, though they do not fully capture the actual vehicle behavior. At the same time, approaches that incorporate full vehicle dynamics have also been widely investigated, providing a more realistic representation of the vehicle motion and enabling higher tracking accuracy \cite{A7,A8,A9}. Many of these studies employ nonlinear control techniques, such as feedback linearization or sliding mode control, to guarantee asymptotic convergence to the desired path and maintain favorable tracking performance under nonlinear vehicle dynamics. However, a significant limitation in these conventional studies \cite{A7,A8,A9} is that they typically treat the steering angle as a direct control input and do not explicitly consider the influence of steering torque resistance arising from the steering mechanism. In such formulations, the effect of torque transmission between the steering actuator and the tires is neglected, which may lead to performance degradation under varying road conditions.

In particular, the stability of path following is strongly influenced by the vehicle's cornering characteristics. When the curvature of the vehicle trajectory converges to the reference value, internal dynamics such as the sideslip angle and the yaw rate---the zero dynamics of the path following system---remain, and their stability must be ensured for stable path following. In this paper, the zero dynamics are formally derived and analyzed in Section~\ref{sec2-zerodynamics}, where it is shown that they are stable whenever the rear cornering power is positive, providing a theoretical foundation for stable path following control.

Even when zero dynamics are stable, variations in vehicle parameters or road conditions may still degrade tracking accuracy. To address such uncertainties, robust control frameworks that explicitly consider model uncertainties have attracted increasing attention. Among them, Model Error Compensator (MEC) provides a systematic approach to enhancing robustness by compensating for discrepancies between the nominal model and the actual system behavior.

MEC is a compensator designed to improve robustness in control systems by feeding back the discrepancy between the actual system output and the model output through an error compensator, thereby mitigating the effects of model uncertainties. Owing to its compatibility with nonlinear systems and its ability to support feedback linearization, MEC is well suited for path following applications \cite{A10,A11,A12}. Previous studies have shown that MEC effectively suppresses tracking errors caused by variations in cornering power arising from differences in road surface conditions \cite{A10}. However, while the MEC framework was applied in \cite{A10} to compensate for uncertainties such as variations in cornering power, the uncertainty associated with steering torque resistance was not explicitly addressed. This limitation becomes critical in environments where steering resistance significantly varies, such as unstructured or unpaved terrains.

In this study, MEC is not merely introduced as an auxiliary compensator but is employed as a fundamental framework for achieving robust path following control in the presence of model uncertainties and external disturbances. Several alternative robust control approaches have been proposed in the literature and are briefly reviewed here to clarify the positioning of the proposed method. Sliding mode control (SMC) achieves robustness through discontinuous control actions that enforce sliding on a predefined manifold; however, it typically suffers from chattering and may require knowledge of disturbance bounds \cite{A13}. Active Disturbance Rejection Control (ADRC), which employs an Extended State Observer (ESO) to estimate and cancel lumped disturbances in real time, has also been widely applied to vehicle control problems \cite{A14}. While ADRC/ESO provides effective disturbance suppression, it is typically formulated for systems expressed in an integrator-chain (canonical) form, and the lumped disturbance is commonly assumed to act in the input channel; its application to systems outside this class requires additional reformulation. Disturbance Observer-Based Control (DOBC) is another widely used approach \cite{A15,A16}. While recent nonlinear extensions of DOBC have relaxed some classical requirements such as the explicit use of a plant inverse, they generally still rely on assumptions regarding the disturbance structure. In contrast, MEC estimates the aggregate model error by comparing the outputs of the actual plant and a nominal model running in parallel, without requiring an explicit plant inverse or a transformation of the system into a specific canonical form. Moreover, MEC is designed as an add-on compensator that can be integrated into an existing control system without modifying the primary controller structure. This modularity makes MEC particularly suitable for the problem considered in this study, where the baseline controller is designed via feedback linearization and the steering torque resistance is treated as an uncertain parameter.

In addition to cornering power, steering torque resistance is another important factor influenced by road conditions. Steering torque resistance refers to the torque required to rotate the steering wheel. In vehicle steering systems, the steering wheel is mechanically connected to the tires, and the resistance generated during steering is influenced by several factors, including road surface conditions and vehicle speed. Although the detailed tire–road interaction is inherently complex, its dominant influence on steering behavior can be captured through simplified modeling. In this study, the steering torque resistance is represented by a simplified model that reflects its dependence on vehicle speed and steering angle, allowing the essential effects of road conditions to be incorporated while maintaining analytical tractability.

The issue of uncertain steering torque is relevant to a broad class of vehicles, including those operating on unstructured or unpaved surfaces. One representative example is agricultural machinery, which is frequently exposed to varying surface conditions that affect steering behavior and often lacks Electric Power Steering (EPS), making it more sensitive to environmental variations \cite{A17}. While the present study focuses on a general vehicle model and validates the proposed method through numerical simulations under idealized conditions, the framework is intended to serve as a foundation for such demanding applications. These considerations motivate the development of path following control strategies that are robust to variations in steering torque resistance.

The main contributions of this study are summarized as follows:
\begin{itemize}
    \item \textbf{Contribution 1:} Development of a vehicle model incorporating steering torque resistance and design of a path following control system based on this model. Unlike the previous study \cite{A10}, where the steering angle was treated as a direct control input, this study models the steering angle as a state variable and treats the torque as the control input.
    
    \item \textbf{Contribution 2:} Formal derivation and stability analysis of the zero dynamics associated with the path following formulation, showing that they are stable for any positive rear cornering power (Section~\ref{sec2-zerodynamics}).
    
    \item \textbf{Contribution 3:} Application of the MEC framework to compensate for uncertainties in steering torque, achieving robust path following performance under varying road surface conditions.
    
    \item \textbf{Contribution 4:} Characterization of the allowable range of parameter mismatch $C/C_{M}$ within which the proposed method maintains practical tracking performance, as demonstrated through systematic numerical evaluation (Table\,\ref{table:robustness1} and Table\,\ref{table:robustness2}).
\end{itemize}

Therefore, this study proposes a robust path following control strategy that incorporates steering torque dynamics into the vehicle model and utilizes the MEC framework to compensate for model uncertainties. By establishing the stability of the zero dynamics associated with path following and by treating the steering torque resistance as an uncertain parameter, the proposed method improves the robustness and accuracy of autonomous vehicle path following under varying road conditions. The effectiveness of the proposed approach is demonstrated through numerical simulations, showing a significant reduction in tracking error compared to conventional methods.

\section{Geometric relationship between vehicle and target path}\label{sec2}
\subsection{Problem formulation}

\begin{figure}[b]
\centering
\includegraphics[width=8.0cm, height=5cm]{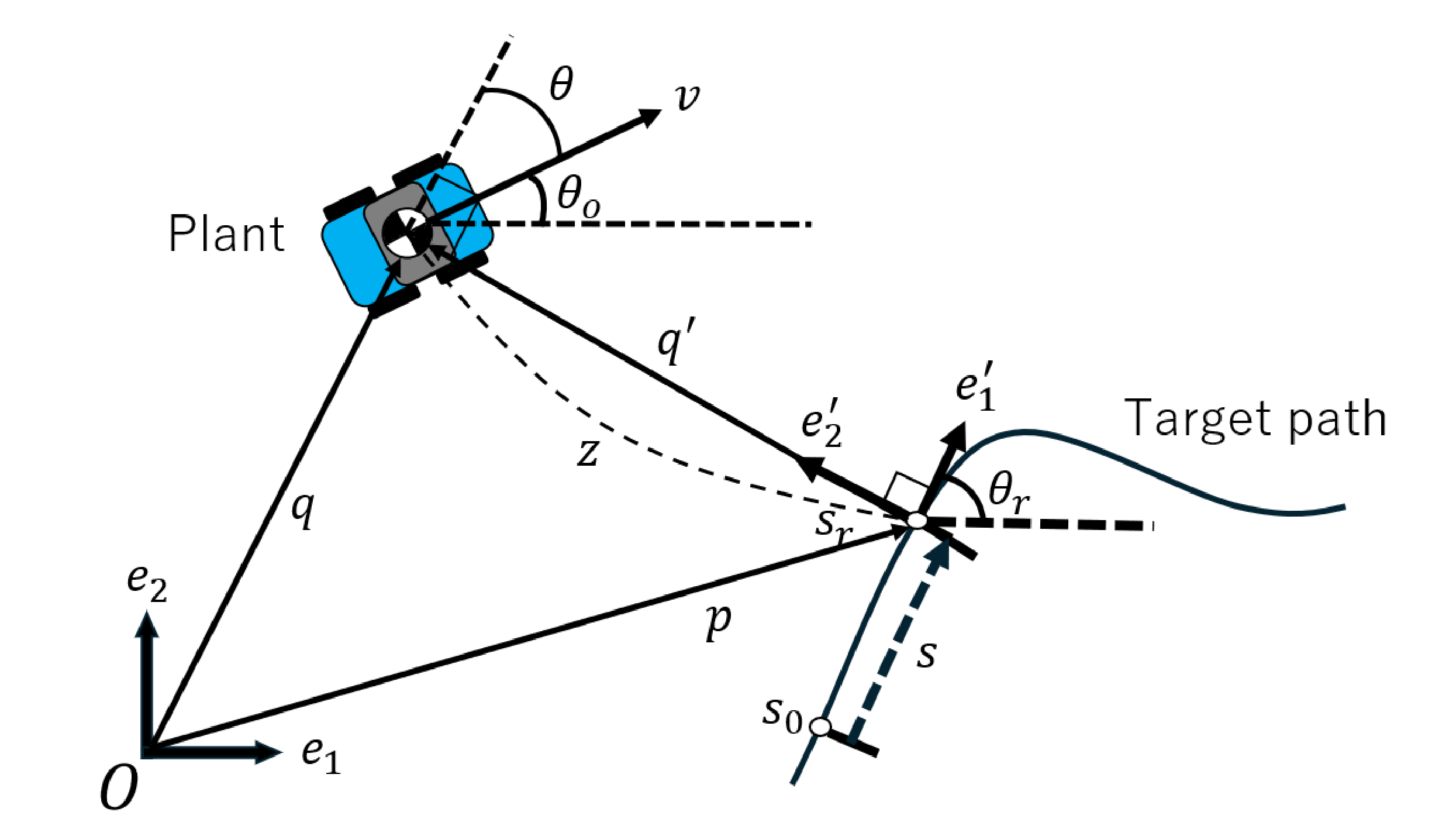}
\caption{Relationship between target path and plant}
\label{Relationship between target path and plant}
\end{figure}

In this study, the target path is defined on a two-dimensional plane, assuming no elevation changes or gradients. Let $\Sigma$ be the inertial coordinate system with an orthonormal basis $\{e_{1}$, $e_{2}\}$. Hereafter, Fig.\,\ref{Relationship between target path and plant} is referenced. A point $p$ on the target path is parameterized by the arc length $s$ measured from a reference point and can be expressed as
\begin{equation}
p(s)=\xi_{r}(s)e_{1}+\eta_{r}(s)e_{2}
\end{equation}
To describe the local geometry of the path, we introduce a moving coordinate system $\Sigma'$ centered at $p(s)$, with an orthonormal basis $\{e'_{1}(s)$, $e'_{2}(s)\}$. Here, $e'_{1}(s)$ is the unit tangent vector in the direction of increasing $s$. This vector is given by
\begin{equation}
\label{2}
e'_{1}(s)=\frac{dp(s)}{ds}=\frac{d\xi_{r}(s)}{ds}e_{1}+\frac{d\eta_{r}(s)}{ds}e_{2}
\end{equation}

Let $\theta_{r}(s)$ denote the orientation of the tangent vector $e'_{1}(s)$ relative to $e_{1}$. Using this definition $e'_{1}(s)$ and $e'_{2}(s)$ can be rewritten as
\begin{align}
\label{3}
e'_{1}(s)&=\cos(\theta_{r}(s))e_{1}+\sin(\theta_{r}(s))e_{2}\\
\label{4}
e'_{2}(s)&=-\sin(\theta_{r}(s))e_{1}+\cos(\theta_{r}(s))e_{2}
\end{align}
The curvature $\kappa_{r}(s)$ of the target path at $p(s)$ is defined as the rate of change of $\theta_{r}(s)$ with respect to $s$:
\begin{equation}
\label{5}
\kappa_{r}(s)=\frac{d\theta_{r}(s)}{ds}
\end{equation}
Since $\xi_{r}$, $\eta_{r}$ are of class $C^3$, the curvature $\kappa_{r}$ is differentiable at each point $s$.

The motion of the plant is assumed to be represented by the following state equation.
\begin{equation}
\dot{x}_{p}=f_{p}(x_{p}(t),u(t))
\label{6}
\end{equation}

Here, $x_{p} \in {\bf{R}}^{n}$ represents the state, and $u(t)$ is the control input. It is assumed that the curvature $\kappa$ of the trajectory traced by the plant can be described as follows.
\begin{align}
\label{7}
v(t)&=h_{1}(x_{p}(t),u(t))\\
\label{8}
\kappa(t)&=h_{2}(x_{p}(t),u(t))
\end{align}
In this case, let the coordinates of the center of mass of the plant be given by
\begin{equation}
q(t)=\xi(t)e_{1}+\eta(t)e_{2}\label{9}
\end{equation}
Then, the components of $q(t)$ can be represented in terms of $\kappa(t)$ and $v(t)$ as follows.
\begin{align}
\xi(t)&=\int_{t_{0}}^{t} v(\tau)\cos\theta_{o}(\tau)d\tau+\xi(t_{0})\label{11}\\
\eta(t)&=\int_{t_{0}}^{t} v(\tau)\sin\theta_{o}(\tau)d\tau+\eta(t_{0})\label{12}\\
\theta_{o}(t)&=\int_{t_{0}}^{t} \kappa(\tau)v(\tau)d\tau+\theta_{o}(t_{0})\label{13}
\end{align}
Here, $\theta_{o}$ represents the angle between the velocity vector and $e_{1}$.

\subsection{Relationship between the target path and the plant}
In this paper, the deviation between the vehicle and the target path is described by means of a reference point on the path \cite{A18,A19}. When a point on the target path changes with time $t$, its coordinates are given by $p_{r}(s_{r}(t))$, where $s_{r}$ is assumed to be differentiable with respect to $t$. For simplicity in notation, we define $p_{r}(s_{r}(t))$ as $\hat{p}_{r}(t)$. Similarly, for other functions of $s_{r}$, when using $s_{r}(t)$ as the argument, we adopt the notation ($\hat{\ }$) to omit $s_{r}$.

Let $q'(t)$ be the vector from $\hat{p}_{r}(t)$ to the center of gravity of the plant. Then, the following relationship holds:
\begin{equation}
q(t)=q'(t)+\hat{p}_{r}(t)
\label{14}
\end{equation}
To quantify the deviation, we define a signed distance related to the length of $q'(t)$ as follows:
\begin{equation}
\zeta(q',s_{r})=\sgn(q'\cdot e'_{2}(s_{r}))\|q'\|
\end{equation}
Here, $\|\cdot\|$ represents the Euclidean norm. The magnitude of $\zeta$ corresponds to the magnitude of $q'$, while the sign of $\zeta$ indicates on which side of the path the plant is located: $\zeta$ is positive when the plant lies in the region where $e'_{2}$ is positive.

\subsection{The state equation related to geometric relationship}
The reference point is chosen such that $q'$ is orthogonal to the tangent direction of the path\cite{A18,A19}. That is, when there exists a differentiable $s_{r}(t)$ satisfying the condition
\begin{equation}
q'(t)\cdot e'_{1}(s_{r}(t))=0
\label{24}
\end{equation}
we define $\hat{p}_{r}(t)$ as the reference point. Such an $s_{r}(t)$ does not necessarily exist for an arbitrary relationship between the trajectory of the plant and the target path; necessary and sufficient conditions for its existence are established in \cite{A18,A19} and are summarized as follows. Suppose that $\kappa(t)$ and $v(t)$ are continuous on $[t_{0},t_{1}]$ and that the orthogonality condition (\ref{24}) holds at the initial time with $1-\hat{\kappa}_{r0}\zeta(q'(t_{0}),s_{r0})>0$. Then, a differentiable $s_{r}(t)$ satisfying (\ref{24}) together with $1-\hat{\kappa}_{r}(t)\zeta(q'(t),s_{r}(t))>0$ exists on $[t_{0},t_{1}]$ if and only if the state equation obtained by differentiating (\ref{14}) with respect to time and using the Frenet relations (\ref{3})--(\ref{5}),
\begin{align}
\label{25}
\dot{x}_{re}&=f_{re}(x_{re},\kappa(t),\hat{\kappa}_{r}(t),v(t))\\
x_{re}&=
\left[
\begin{array}{ccc}
\theta & s_{r} & z
\end{array}
\right]^{\rm{T}}\nonumber\\
f_{re}&=
\left[
\begin{array}{c}
\kappa v-\hat{\kappa}_{r}\dfrac{v\cos\theta}{1-\hat{\kappa}_{r}z}\\
\dfrac{v\cos\theta}{1-\hat{\kappa}_{r}z}\\
v\sin\theta
\end{array}
\right]\nonumber\\
\theta(t_{0})&=\theta_{o}(t_{0})-\hat{\theta}_{r}(t_{0})\\
s_{r}(t_{0})&=s_{r0}\\
z(t_{0})&=\zeta(q'(t_{0}),s_{r0})
\end{align}
has a bounded solution on $[t_{0},t_{1}]$ \cite{A18,A19}. Here, $\theta=\theta_{o}-\hat{\theta}_{r}$ is the relative angle between the direction of the vehicle velocity and the tangent of the path, and $z=\zeta$ represents the signed distance between the plant and the reference point. Throughout this paper, we assume that these existence conditions are satisfied.

The state equation (\ref{25}) is the basis of the subsequent control design: the geometric states $\theta$, $s_{r}$, and $z$ are directly used in Section~\ref{sec3} to formulate the path following control law that ensures $z \to 0$, with $z$ serving as the primary tracking error to be regulated.

\subsection{Model error occurring in the vehicle}
\begin{figure}[t]
\centering
\includegraphics[width=6.5cm, height=6cm]{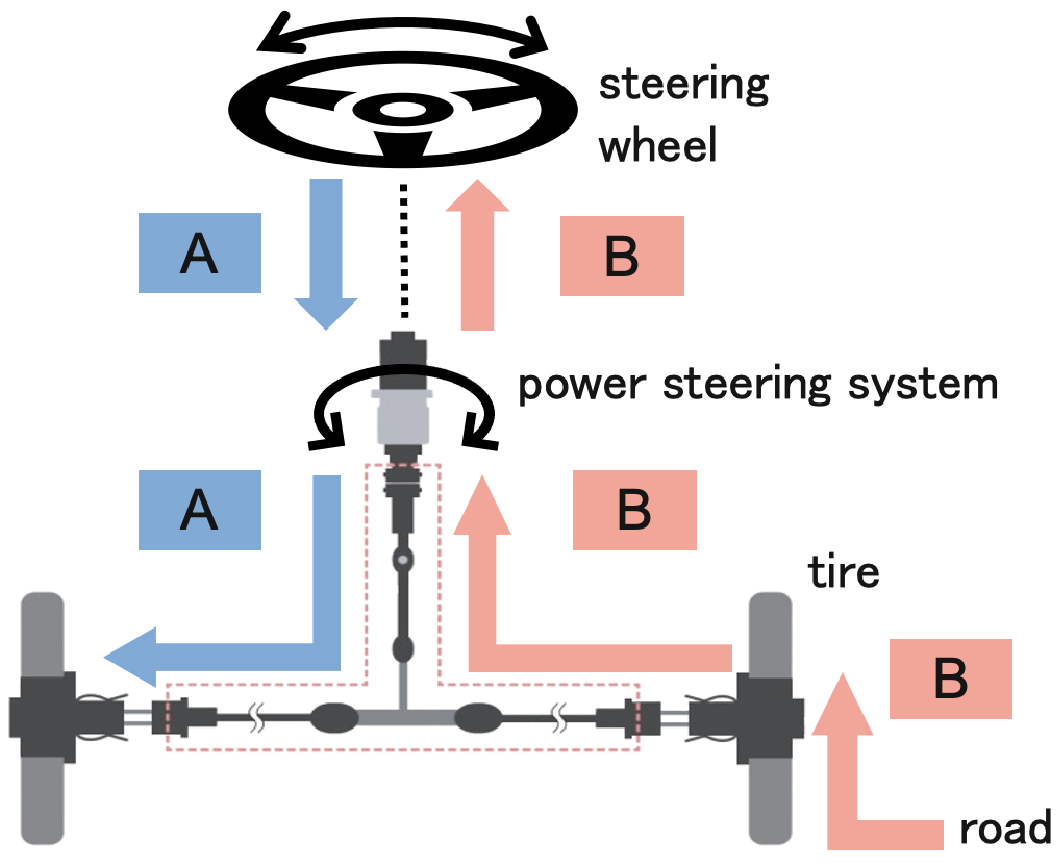}
\caption{Steering system}
\label{Steering system}
\end{figure}

Here, we describe the steering dynamics and the model error considered in this study. In \cite{A10}, the steering angle $\delta$ was treated as a direct control input, meaning that the steering wheel angle directly determined the steering angle (case A in Fig.\,\ref{Steering system}). In practice, however, resistance to the steering input is introduced from the road surface to the steering wheel through the tires and the steering mechanism (case B in Fig.\,\ref{Steering system}). The magnitude of this resistance is influenced by road surface irregularities, vehicle speed, and vehicle weight\cite{A20}; in addition, the self-aligning torque, which acts to return the tire to the straight-ahead direction when slip occurs, also contributes to the resistance and increases with the steering angle. This study therefore incorporates the steering angle as a state variable driven by the steering command $u$ against the resistance term $Cv\delta$, resulting in the first-order dynamics
\begin{equation}
\dot{\delta}=u-Cv\delta
\label{eq:steerdyn}
\end{equation}
Here, $u$ denotes the steering command [rad/s], and $Cv\delta$ represents the steering torque resistance term, where $C$ [1/m] is a coefficient that varies depending on road conditions. This first-order model assumes that the inertia of the steering mechanism is sufficiently small relative to the resistance and input magnitudes, which is particularly reasonable for low-speed applications such as agricultural vehicles. On asphalt roads or dry surfaces, the value of $C$ is relatively small, while on farmlands or muddy surfaces, it increases. Since $C$ depends on the environment and is difficult to determine accurately under real driving conditions, a constant nominal value assumed in the model inevitably deviates from the actual value, and the resulting model error becomes a significant factor affecting the path following performance of the vehicle.
Note that the first-order model (\ref{eq:steerdyn}) can be derived from the torque balance of the steering mechanism. Let $J$ denote the inertia of the steering mechanism, $b\dot{\delta}$ the viscous resistance torque, and $kv\delta$ the road-reaction torque, which is dominated by the self-aligning torque and grows with the vehicle speed and the steering angle; the torque balance is then $J\ddot{\delta}=\tau-b\dot{\delta}-kv\delta$, where $\tau$ denotes the steering torque applied to the steering mechanism. When the inertia $J$ is sufficiently small relative to the viscous and reaction torques, the quasi-static approximation yields $b\dot{\delta}\approx\tau-kv\delta$, that is, $\dot{\delta}=u-Cv\delta$ with $u=\tau/b$ [rad/s] and $C=k/b$ [1/m], which coincides with (\ref{eq:steerdyn}). Thus $u$ is proportional to the applied steering torque, and the coefficient $C$ inherits the road-surface dependence of the reaction torque; accordingly, the term ``steering torque'' in the title of this paper refers to this torque-level uncertainty, while $u$ itself has the dimension of a steering rate. Throughout this paper, $u$ is referred to as the steering command.

The assumption that the steering torque resistance is proportional to both the speed and the steering angle is not merely a mathematical simplification, but is grounded in physical behavior observed in real-world vehicle dynamics. In actual driving conditions, as the steering angle $\delta$ increases, tire--road interactions become more significant, resulting in an increase in steering torque resistance \cite{A21}. Moreover, as the vehicle speed $v$ increases, the lateral forces acting on the tires grow, which in turn amplifies the reaction torque to steering input \cite{A22,A23}. Therefore, the steering torque resistance can be regarded as a physically motivated model reflecting these phenomena, and is expressed in the proportional form of $Cv\delta$.
\subsection{Vehicle dynamics}
\begin{figure}[b]
\centering
\includegraphics[width=8.0cm, height=5.0cm]{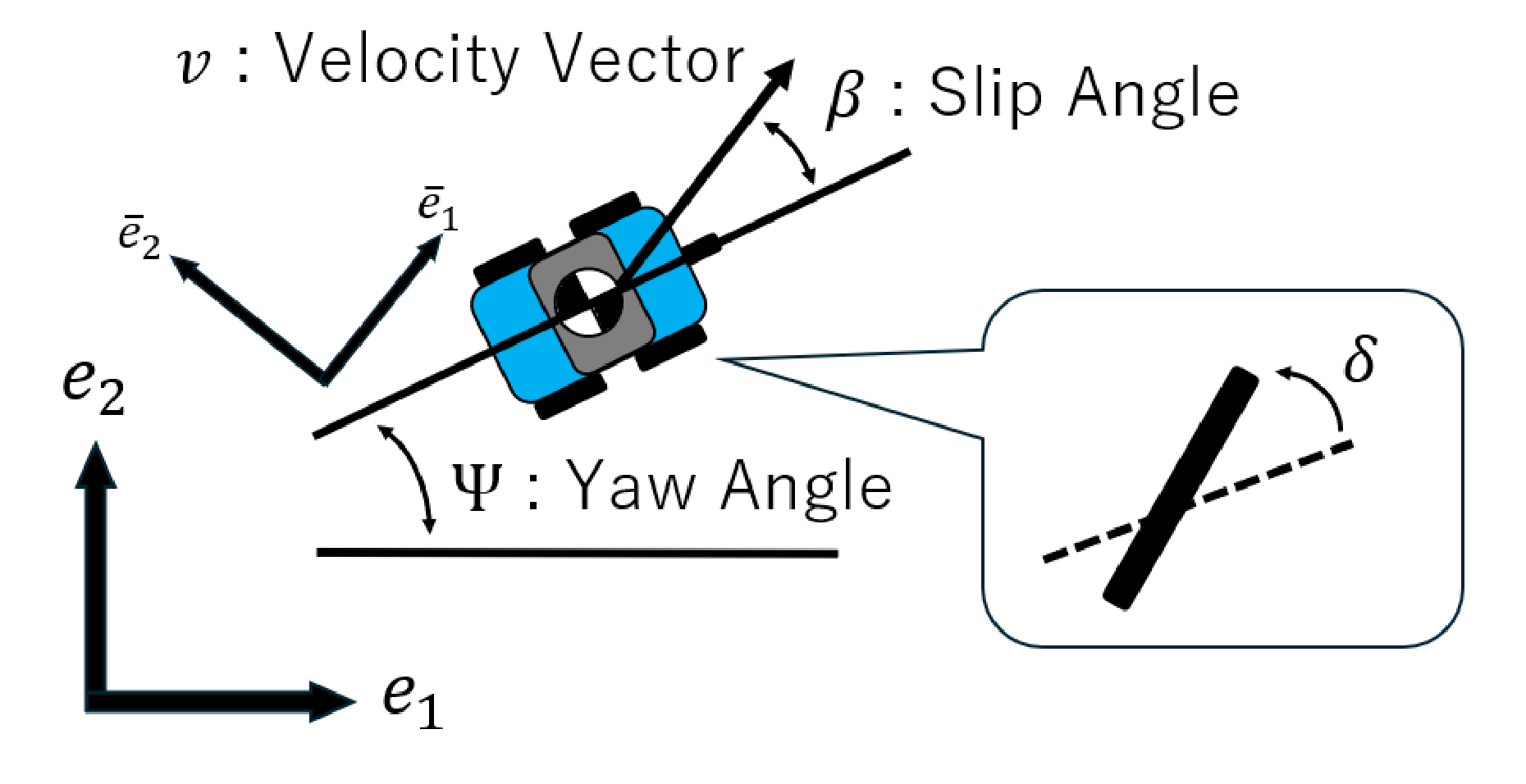}
\caption{Vehicle model}
\label{Vehicle model}
\end{figure}

In this study, we consider a four-wheeled vehicle as the plant. The vehicle model, shown in Fig.\,\ref{Vehicle model}, is represented using an equation of motion where the yaw rate $\dot{\psi}$, sideslip angle $\beta$, and front wheel steering angle $\delta$ are treated as state variables, while the vehicle speed is assumed to be constant. The steering command $u$ is used as the control input. The motion model is expressed in a coordinate system ($\bar{e}_{1}$,$\bar{e}_{2}$) aligned with the vehicle velocity vector. In this coordinate system, the vehicle velocity vector and the forces acting on the center of mass of the vehicle are given by
\begin{align}
v&=v_{\bar{e}_{1}}(t)\bar{e}_{1}(t)+v_{\bar{e}_{2}}(t)\bar{e}_{2}(t)\\
f&=f_{\bar{e}_{1}}(t)\bar{e}_{1}(t)+f_{\bar{e}_{2}}(t)\bar{e}_{2}(t)
\end{align}
Furthermore, by considering the rotation of the coordinate system, the following equations of the motion are obtained:
\begin{align}
m(\dot{v}_{\bar{e}_{1}}(t)-v_{\bar{e}_{2}}(t)(\dot{\psi}(t)+\dot{\beta}(t)))&=f_{\bar{e}_{1}}(t)\\
m(\dot{v}_{\bar{e}_{2}}(t)+v_{\bar{e}_{1}}(t)(\dot{\psi}(t)+\dot{\beta}(t)))&=f_{\bar{e}_{2}}(t)\\
I\ddot{\psi}(t)&=M(t)
\end{align}
Here, $m$ represents the mass of the vehicle, $I$ denotes the moment of inertia about the center of the mass of the vehicle, and $M$ is the total torque acting around the center of the mass of the vehicle. Given the definitions of $\bar{e}_{1}$, $\bar{e}_{2}$ and considering the vehicle speed $v(t)$, the velocity components satisfy $v_{\bar{e}_1}(t)=v$ and $v_{\bar{e}_2}(t)=0$. The forces $f$ and torque $M$ are determined based on the assumption that lateral forces are generated in proportion to the slip angles of the tires. Since the slip angles can be approximately expressed using the sideslip angle of the vehicle $\beta$ and the distance between the axles, the following state equations hold \cite{A10}.
\begin{align}
\dot{\beta}&=\frac{a_{11}}{v}\beta+\Bigg(-1+\frac{a_{12}}{v^2}\Bigg)\dot{\psi}+\frac{a_{13}}{v}\delta\\
\ddot{\psi}&=a_{21}\beta+\frac{a_{22}}{v}\dot{\psi}+a_{23}\delta
\end{align}
Here, $a_{ij}$ denotes coefficients determined by the vehicle parameters, defined as follows \cite{A10}:
\begin{align}
&a_{11}=-\frac{K_{f}+K_{r}}{m},\quad a_{12}=-\frac{l_{f}K_{f}-l_{r}K_{r}}{m}, \nonumber\\
&a_{13}=\frac{K_{f}}{m},\quad a_{21}=-\frac{l_{f}K_{f}-l_{r}K_{r}}{I},\label{eq:aij2}\\
&a_{22}=-\frac{l_{f}^{2}K_{f}+l_{r}^{2}K_{r}}{I},\quad a_{23}=\frac{l_{f}K_{f}}{I}\nonumber
\end{align}
where $K_{f}$ and $K_{r}$ denote the cornering power of the front and rear tires, respectively, $l_{f}$ and $l_{r}$ are the distances from the center of gravity to the front and rear axles, respectively, and $I$ is the moment of inertia about the vertical axis through the center of gravity.

Based on the above considerations, the following state equations are obtained, where the steering command $u$ serves as the control input, and the sideslip angle $\beta$, vehicle yaw rate $\dot{\psi}$, and steering angle $\delta$ are treated as state variables. In this study, the velocity is assumed to be constant.
\begin{align}
\dot{x}_{p}&=f_{p}(x_{p}(t),u(t))\\
x_{p}&=
\left[
\begin{array}{cccc}
\beta & \dot{\psi} & \delta
\end{array}
\right]^{\rm{T}}\nonumber\\
f_{p}(x_{p},u)&=
\left[
\begin{array}{c}
\dfrac{a_{11}}{v}\beta+\Bigg(-1+\dfrac{a_{12}}{v^2}\Bigg)\dot{\psi}+\dfrac{a_{13}}{v}\delta\\
a_{21}\beta+\dfrac{a_{22}}{v}\dot{\psi}+a_{23}\delta\\
u-Cv\delta\\
\end{array}
\right]\nonumber
\end{align}

At this time, the curvature of the trajectory followed by the plant is given by the following equation.
\begin{equation}
h_{2}(x_{p},u)=\frac{a_{11}}{v(t)^2}\beta(t)+\frac{a_{12}}{v(t)^3}\dot{\psi}(t)+\frac{a_{13}}{v(t)^2}\delta(t)
\label{eq:h2curv}
\end{equation}

\subsection{Zero dynamics and cornering power}\label{sec2-zerodynamics}
In the control design of Section~\ref{sec3}, the control law drives the tracking error $z$ to zero. When $z\equiv0$ (and consequently $\theta\equiv0$), the vehicle trajectory coincides with the target path, but the sideslip angle $\beta$ and yaw rate $\dot{\psi}$ continue to evolve. These remaining internal dynamics, called the \emph{zero dynamics}, must be stable for the closed-loop system to function properly. In this subsection, we derive the zero dynamics and show that their stability is guaranteed by a simple condition on the cornering power.

Under $z\equiv0$, the vehicle curvature matches the path curvature, $\kappa=\hat{\kappa}_{r}$. From (\ref{eq:h2curv}), this constrains the steering angle to
\begin{equation}
\delta=\frac{1}{a_{13}}\Big(\hat{\kappa}_{r}v^{2}-a_{11}\beta-\frac{a_{12}}{v}\dot{\psi}\Big)
\label{eq:zerodelta}
\end{equation}
Substituting (\ref{eq:zerodelta}) into the $\beta$--$\dot{\psi}$ dynamics eliminates $\delta$ and yields the zero dynamics
\begin{align}
\left[
\begin{array}{c}
\dot{\beta}\\
\ddot{\psi}
\end{array}
\right]
&=A_{\zeta}
\left[
\begin{array}{c}
\beta\\
\dot{\psi}
\end{array}
\right]
+b_{\zeta}\hat{\kappa}_{r}
\label{eq:zerodyn}\\
A_{\zeta}&=
\left[
\begin{array}{cc}
0 & -1\\
a_{21}-\dfrac{a_{23}a_{11}}{a_{13}} & \dfrac{1}{v}\Bigg(a_{22}-\dfrac{a_{23}a_{12}}{a_{13}}\Bigg)
\end{array}
\right],\nonumber\\
b_{\zeta}&=
\left[
\begin{array}{c}
v\\
\dfrac{a_{23}}{a_{13}}v^{2}
\end{array}
\right]\nonumber
\end{align}
Since $v$ is constant, $A_{\zeta}$ is a constant matrix. The zero dynamics (\ref{eq:zerodyn}) are exponentially stable if and only if $A_{\zeta}$ is Hurwitz, which requires
\begin{equation}
a_{21}-\frac{a_{23}a_{11}}{a_{13}}>0,\quad a_{22}-\frac{a_{23}a_{12}}{a_{13}}<0
\label{eq:zerostab}
\end{equation}
Substituting the definitions of $a_{ij}$ from (\ref{eq:aij2}) into (\ref{eq:zerostab}) gives
\begin{equation}
a_{21}-\frac{a_{23}a_{11}}{a_{13}}=\frac{K_{r}l}{I},\quad
a_{22}-\frac{a_{23}a_{12}}{a_{13}}=-\frac{K_{r}l_{r}l}{I}
\label{eq:zerophys}
\end{equation}
where $l=l_{f}+l_{r}$ is the wheelbase. Since $K_{r}$, $l$, $l_{r}$, and $I$ are all positive physical quantities, both inequalities in (\ref{eq:zerostab}) are automatically satisfied whenever the rear cornering power $K_{r}$ is positive---a condition that holds for any pneumatic tire under normal operating conditions. Therefore, for the linearized bicycle model employed in this study, the zero dynamics are stable without any additional condition beyond $K_{r}>0$, providing the theoretical foundation for the path following control design in Section~\ref{sec3}.

For the vehicle parameters used in Section~\ref{sec4}, the eigenvalues of $A_{\zeta}$ are $\lambda_{1}\approx-2.59$ and $\lambda_{2}\approx-33.3$, confirming rapid decay of the internal dynamics.

\begin{remark}
The understeer condition $K_{f}l_{f}/(K_{r}l_{r})<1$ is a well-known steady-state cornering property of conventional passenger vehicles \cite{A7,A9} and is satisfied by the simulation parameters ($K_{f}l_{f}/(K_{r}l_{r})\approx0.63$). However, the stability of the zero dynamics is governed by (\ref{eq:zerostab}), not by the understeer condition; (\ref{eq:zerostab}) holds for both understeer and oversteer vehicles as long as $K_{r}>0$.
\end{remark}

\begin{remark}
In the subsequent Section~\ref{sec3}, the control target is the combined system $P_\sigma$ consisting of the steering dynamics $P_d$ given by (\ref{eq:steerdyn}) and the vehicle plant $P$ (with states $\beta$, $\dot{\psi}$, $\delta$), together with the geometric state equation (\ref{25}) for $\theta$, $s_r$, and $z$. The control input is the steering torque command $u$, and the objective is to design $u$ such that the tracking error $z$ converges to zero.
\end{remark}

\section{Robust path following control} \label{sec3}
\subsection{Control law for tracking the target path} \label{sec3-1}
If the condition $z=0$ can be maintained at all times during operation, it is expected that the vehicle will follow the target path accurately. The steering control law achieves this by imposing the constraint derived in \cite{A9}, which ensures that the tracking error $z$ asymptotically converges to zero. In \cite{A9}, the condition
\begin{equation}
\lim_{t\to \infty} z = 0
\label{3.1}
\end{equation}
is considered as a criterion for achieving path following. In this section, we first discuss the control law that satisfies (\ref{3.1}) under the assumption that the steering resistance is zero ($C=0$).

For coefficients $\alpha_{1}$,$\alpha_{2}$, $\alpha_{3}$ that ensure $s^3+\alpha_{1}s^2+\alpha_{2}s+\alpha_{3}$ is a Hurwitz polynomial, the condition
\begin{equation}
\frac{d^3z}{dt^3}+\alpha_{1}\frac{d^2z}{dt^2}+\alpha_{2}\frac{dz}{dt}+\alpha_{3}z=0
\label{3.2}
\end{equation}
must hold to satisfy (\ref{3.1}). Here, $d^2z/dt^2$ is given by
\begin{equation}
\frac{d^2z}{dt^2}=\cos\theta\Bigg(a_{11}\beta+\frac{a_{12}}{v}\dot{\psi}+a_{13}\delta-\kappa_{r}\frac{v^2\cos\theta}{1-\kappa_{r}z}\Bigg)
\label{3.3}
\end{equation}
Additionally, $d^3z/dt^3$ is expressed as
\begin{equation}
\begin{split}
\frac{d^3z}{dt^3}=&\Bigg(a_{11}\dot{\beta}+\frac{a_{12}}{v}\ddot{\psi}+a_{13}\dot{\delta}\Bigg)\cos\theta\\-&\frac{1}{v}\Bigg(a_{11}\beta+\frac{a_{12}}{v}\dot{\psi}+a_{13}\delta\Bigg)^2\sin\theta\\+&\frac{3\kappa_{r}v\sin\theta\cos\theta}{1-\kappa_{r}z}\Bigg(a_{11}\beta+\frac{a_{12}}{v}\dot{\psi}+a_{13}\delta\Bigg)\\-&\frac{\dot{\kappa}_{r}v^2\cos^2\theta}{1-\kappa_{r}z}-\frac{\kappa_{r}\dot{\kappa}_{r}v^2z\cos^2\theta}{(1-\kappa_{r}z)^2}\\-&\frac{3\kappa_{r}^2v^3\sin\theta\cos^2\theta}{(1-\kappa_{r}z)^2}
\label{3.4}
\end{split}
\end{equation}
Therefore, setting $C=0$ (i.e., $\dot{\delta}=u$) in the above expressions, the ideal steering command $u_{0}$ in the absence of steering resistance is given by
\begin{equation}
\begin{split}
&u_{0}=\\&-\alpha_{1}\frac{1}{a_{13}}\Bigg(a_{11}\beta+\frac{a_{12}}{v}\dot{\psi}+a_{13}\delta-\frac{\kappa_{r}v^2\cos\theta}{1-\kappa_{r}z}\Bigg)\\&-\alpha_{2}\frac{v\tan\theta}{a_{13}}-\alpha_{3}\frac{z}{a_{13}\cos\theta}-\frac{1}{a_{13}}\Bigg(a_{11}\dot{\beta}+\frac{a_{12}}{v}\ddot{\psi}\Bigg)\\&+\frac{\tan\theta}{a_{13}v}\Bigg(a_{11}\beta+\frac{a_{12}}{v}\dot{\psi}+a_{13}\delta\Bigg)^2\\&-\frac{3\kappa_{r}v\sin\theta}{a_{13}(1-\kappa_{r}z)}\Bigg(a_{11}\beta+\frac{a_{12}}{v}\dot{\psi}+a_{13}\delta\Bigg)\\&+\frac{\dot{\kappa}_{r}v^2\cos\theta}{a_{13}(1-\kappa_{r}z)}+\frac{\kappa_{r}v^2\cos\theta(\dot{\kappa}_{r}z+3\kappa_{r}v\sin\theta)}{a_{13}(1-\kappa_{r}z)^2}
\label{3.5}
\end{split}
\end{equation}

Since this study considers steering torque, the ideal steering command $u$ that accounts for steering resistance can be derived from (\ref{3.5}) as
\begin{eqnarray}
u=u_{0}+Cv\delta
\label{3.6}
\end{eqnarray}
Here, $u_{0}$ represents the ideal term for following a target path, while $Cv\delta$ is introduced to counteract the influence of unknown steering torque resistance. However, $C$ is generally unknown and varies depending on road conditions. Therefore, (\ref{3.6}) represents an ideal steering command, and in practice, a model error inevitably arises due to the uncertainty in $C$. To achieve stable path-following performance under such uncertainties, it is necessary to suppress the effect of this model error.

\subsection{Model error compensator} \label{sec3-2}
\begin{figure}[b]
\centering
\includegraphics[width = 8cm, height = 5cm]{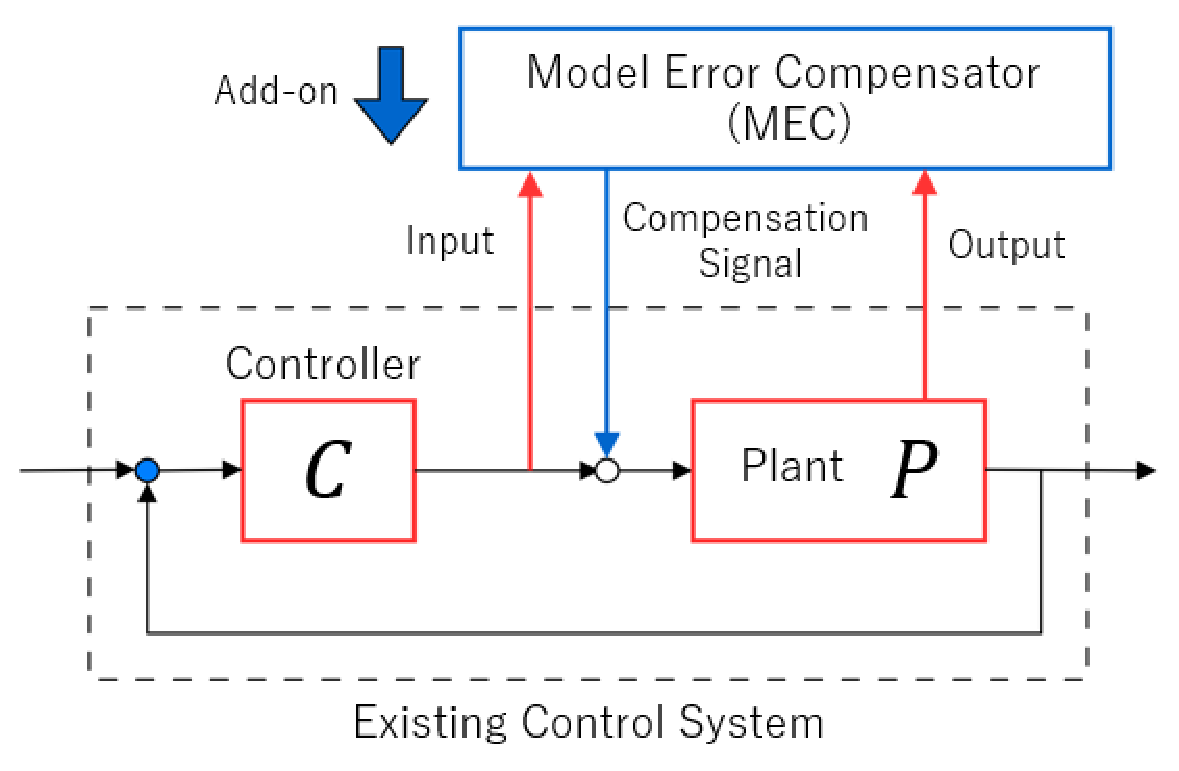}
\caption{Model Error Compensator (MEC)}
\label{mec}
\end{figure}

Model Error Compensator (MEC) is a systematic framework designed to enhance the robustness of existing control systems against model uncertainties and external disturbances \cite{A12}. As illustrated in Fig.\,\ref{mec}, MEC is implemented in an add-on configuration, which enables the integration of the compensator into an existing control system without necessitating a redesign of the primary controller. This modular structure is particularly suitable for systems where the internal parameters of the main control loop are fixed or where modification of the existing control structure is limited by implementation constraints.

The fundamental concept of MEC is to estimate discrepancies—referred to as model errors—between the actual plant and its nominal model based on observed input–output behavior, and to compensate for their influence in real time. Parameter variations, inherent nonlinearities, and unmodeled dynamics are treated as an aggregated error signal, from which a compensation signal is generated and injected into the control input.

Disturbance observer–based control (DOBC) is a well-established approach for enhancing the robustness of control systems \cite{A15}. The classical frequency-domain disturbance observer originally proposed by Kouhei Ohnishi requires the inverse of a nominal plant model, which limits its applicability to nonlinear or non-minimum phase systems \cite{A16}. Although nonlinear extensions such as the approach proposed by Wen-Hua Chen avoid the explicit use of a plant inverse, they typically rely on assumptions regarding the disturbance structure and matching conditions in the disturbance channel.

In contrast, MEC estimates the aggregate model error by comparing the outputs of the actual plant and a nominal model running in parallel, without requiring an explicit plant inverse and without imposing a specific structure on the disturbance channel. Owing to this structural simplicity, MEC is particularly suitable as an add-on compensator for nonlinear systems in which the baseline controller is designed via feedback linearization.

\begin{remark}[Comparison between MEC and ESO]
Both MEC and the Extended State Observer (ESO) aim to enhance robustness against model uncertainties, but they differ in several key aspects.
\begin{itemize}
    \item \textbf{Estimation strategy:} ESO augments the system state with a lumped disturbance term and estimates it using a high-gain observer. MEC estimates the aggregate model error by comparing the outputs of the actual plant and a parallel nominal model, without augmenting the state.
    \item \textbf{Structural requirements:} ESO is typically designed for systems expressed in an integrator-chain (canonical) form, and the lumped disturbance is commonly assumed to enter through the input channel; extensions beyond this class require additional treatment. MEC does not rely on a transformation into such a specific form, since it only compares the measured outputs of the actual plant and the parallel nominal model.
    \item \textbf{Controller integration:} ESO is typically integrated as part of the control loop (within ADRC). MEC is designed as an add-on compensator that can be appended to an existing controller without modifying its internal structure.
    \item \textbf{Applicability:} The add-on nature of MEC is particularly advantageous when the baseline controller (e.g., feedback linearization-based path following) is already designed and its parameters are fixed.
\end{itemize}
\end{remark}

\subsection{Proposed robust path following control law}
\begin{figure}[!t]
\centering
\includegraphics[width = 8.5cm, height = 5cm]{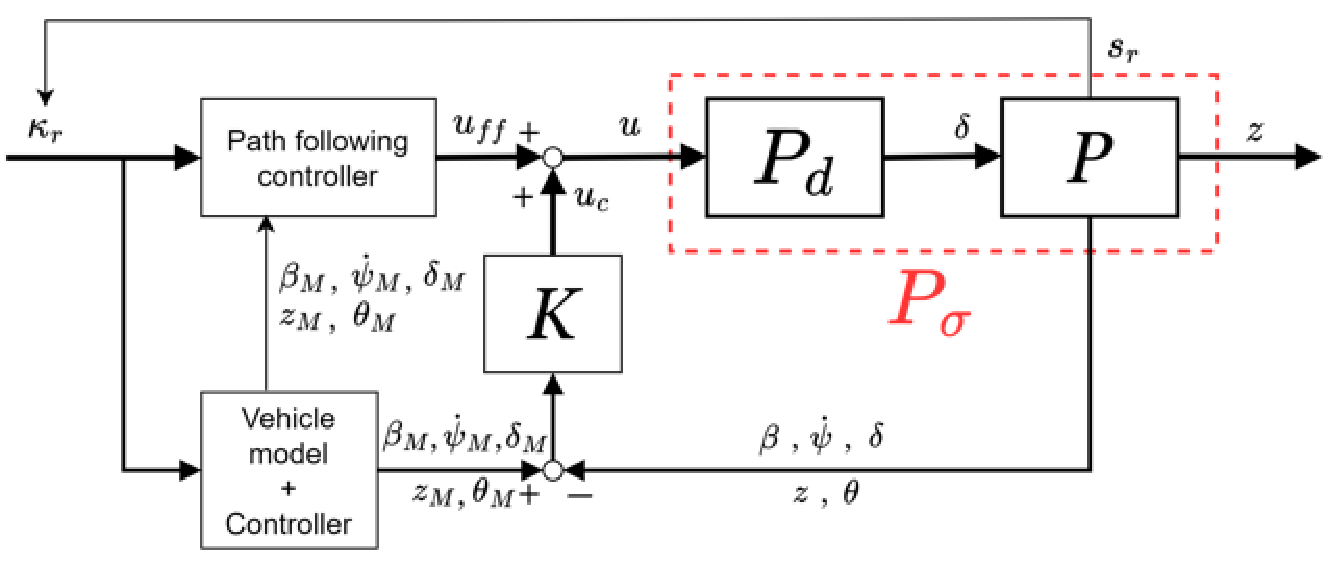}
\caption{Robust path following method based on MEC}
\label{fig1}
\end{figure}

This section describes the robust path following control law, which applies MEC to the path following control law introduced in Section~\ref{sec3-1}. The structure of the control system used in this study is shown in Fig.\,\ref{fig1}. The control system consists of two feedforward controllers, which take the curvature at the reference point as input, and a feedback controller. Here, it is assumed that the relative distance $z$, relative angle $\theta$, and reference point position $s_{r}$ of the plant $P$ can be measured. Additionally, the input is composed of the sum of the feedforward input and the compensation input.

Unlike previous studies where only the plant $P$ was considered as the control target, this study defines the combined dynamics consisting of $P_{d}$ and $P$ as $P_{\sigma}$, and treats this $P_{\sigma}$ as the control target. Here, $P_{d}$ represents a dynamics-like element that models the relationship between the control input and the steering angle $\delta$, governed by $\dot{\delta}=u-Cv\delta$. By incorporating $P_{d}$ into the control target, the dynamic characteristics between the input and the steering angle $\delta$ are explicitly considered, allowing the control design to more accurately reflect the dynamic relationship from the control input to the vehicle response. In this study, values computed from the model on a computer are denoted with the subscript $M$.

The control input is constructed as $u = u_M + u_c$, where $u_M$ is the feedforward input computed from a nominal model running in parallel, and $u_c$ is a feedback compensation input designed to suppress the model error. The derivation of each component is described below.

\subsubsection{Feedforward input $u_M$}
In Fig.\,\ref{fig1}, [Vehicle model\,+\,Controller] represents the application of the control law in (\ref{3.6}) to the nominal vehicle model, designed to satisfy $z_{M}\to0$. Here, the model states are $\beta_{M}$, $\dot{\psi}_{M}$, $\delta_{M}$, $z_{M}$, and $\theta_{M}$, which are computed by integrating the nominal vehicle dynamics with the nominal steering resistance coefficient $C_M$. The feedforward input $u_{M}$ is obtained by substituting these model states into the control law (\ref{3.6}), yielding:
\begin{equation}
\begin{split}
&u_{M}=\\&-\alpha_{1}\frac{1}{a_{13}}\Bigg(a_{11}\beta_{M}+\frac{a_{12}}{v}\dot{\psi}_{M}+a_{13}\delta_{M}-\frac{\kappa_{r}v^2\cos\theta_{M}}{1-\kappa_{r}z_{M}}\Bigg)\\&-\alpha_{2}\frac{v\tan\theta_{M}}{a_{13}}-\alpha_{3}\frac{z_{M}}{a_{13}\cos\theta_{M}}\\&-\frac{1}{a_{13}}\Bigg(a_{11}\dot{\beta}_{M}+\frac{a_{12}}{v}\ddot{\psi}_{M}\Bigg)\\&+\frac{\tan\theta_{M}}{a_{13}v}\Bigg(a_{11}\beta_{M}+\frac{a_{12}}{v}\dot{\psi}_{M}+a_{13}\delta_{M}\Bigg)^2\\&-\frac{3\kappa_{r}v\sin\theta_{M}}{a_{13}(1-\kappa_{r}z_{M})}\Bigg(a_{11}\beta_{M}+\frac{a_{12}}{v}\dot{\psi}_{M}+a_{13}\delta_{M}\Bigg)\\&+\frac{\dot{\kappa}_{r}v^2\cos\theta_{M}}{a_{13}(1-\kappa_{r}z_{M})}+\frac{\kappa_{r}v^2\cos\theta_{M}(\dot{\kappa}_{r}z_{M}+3\kappa_{r}v\sin\theta_{M})}{a_{13}(1-\kappa_{r}z_{M})^2}\\&+C_{M}v\delta_{M}
\label{3.7}
\end{split}
\end{equation}
The outputs of [Vehicle model\,+\,Controller] are $\beta_{M}$, $\dot{\psi}_{M}$, $\delta_{M}$, $z_{M}$, $\theta_{M}$. Furthermore, the upper feedforward input $u_{ff}$ in Fig.\,\ref{fig1} is set as $u_{ff}=u_{M}$, utilizing the model states from [Vehicle model\,+\,Controller]. Note that (\ref{3.7}) has the same structure as the ideal control law (\ref{3.6}), but is evaluated using the model states rather than the actual plant states. When the model perfectly matches the plant (i.e., $C=C_M$ and identical initial conditions), the model states coincide with the plant states, and $u_M$ alone achieves $z \to 0$.

\subsubsection{Feedback compensation input $u_c$}
When the plant parameter $C$ differs from the nominal value $C_M$, a discrepancy arises between the model states and the actual plant states. To minimize $|z-z_{M}|$ caused by such model errors, the feedback controller applies a correction $u_{c}$ based on the difference between the model and actual states. The compensation input is defined as:
\begin{equation}
\begin{split}
&u_{c}=\\&\ k_{1}\frac{1}{a_{13}}\Bigg\{\Bigg(a_{11}\beta_{M}+\frac{a_{12}}{v}\dot{\psi}_{M}+a_{13}\delta_{M}-\frac{\kappa_{r}v^2\cos\theta_{M}}{1-\kappa_{r}z_{M}}\Bigg)\\&-\Bigg(a_{11}\beta+\frac{a_{12}}{v}\dot{\psi}+a_{13}\delta-\frac{\kappa_{r}v^2\cos\theta}{1-\kappa_{r}z}\Bigg)\Bigg\}\\&+k_{2}\Bigg(\frac{v\tan\theta_{M}}{a_{13}}-\frac{v\tan\theta}{a_{13}}\Bigg)\\&+k_{3}\Bigg(\frac{z_{M}}{a_{13}\cos\theta_{M}}-\frac{z}{a_{13}\cos\theta}\Bigg)
\label{3.8}
\end{split}
\end{equation}
Here, $k_{1}$, $k_{2}$, and $k_{3}$ serve as the feedback gains. The structure of $u_c$ is designed to mirror the linearizing control law: $k_1$ corresponds to the gain for the second derivative of $z$ (analogous to $\alpha_1$), $k_2$ to the first derivative (analogous to $\alpha_2$), and $k_3$ to $z$ itself (analogous to $\alpha_3$). This correspondence can be made precise as follows: for small relative angles ($\theta,\theta_{M}\approx0$), the first bracketed difference in (\ref{3.8}) coincides, to first order, with the deviation of $d^{2}z/dt^{2}$ between the parallel model and the actual plant (cf.~(\ref{3.3})), the second bracketed difference with the deviation of $dz/dt$ (since $dz/dt=v\sin\theta\approx v\tan\theta$), and the third with the deviation of $z$ itself. Hence, $u_{c}$ constitutes feedback of the output error $z_{M}-z$ and its first and second derivatives, in direct correspondence with the roles of $\alpha_{1}$, $\alpha_{2}$, and $\alpha_{3}$ in (\ref{3.2}). By feeding back the difference between the model states and the actual plant states through these gains, the compensation input effectively suppresses the tracking error caused by the model mismatch.

\subsubsection{Final control input}
As a result, the final control input is given by
\begin{eqnarray}
u=u_{M}+u_{c}
\label{eq:final_input}
\end{eqnarray}
In the feedback component $K$, feedback is applied to the deviation between model and actual states. There are no constraints on the design of $K$, as long as the compensator ensures the stability of the feedback system. Following the MEC design principle of employing high-gain feedback in the frequency band requiring robustification \cite{A12}, the feedback gains $k_1$, $k_2$, $k_3$ are chosen to be sufficiently larger than the Hurwitz coefficients $\alpha_1$, $\alpha_2$, $\alpha_3$ to ensure fast error correction relative to the nominal convergence rate.

\section{Simulation}\label{sec4}
\subsection{Simulation conditions}
\begin{figure}[t]
\centering
\begin{minipage}[t]{0.49\columnwidth}
\centering
\includegraphics[width=\columnwidth, height=4cm]{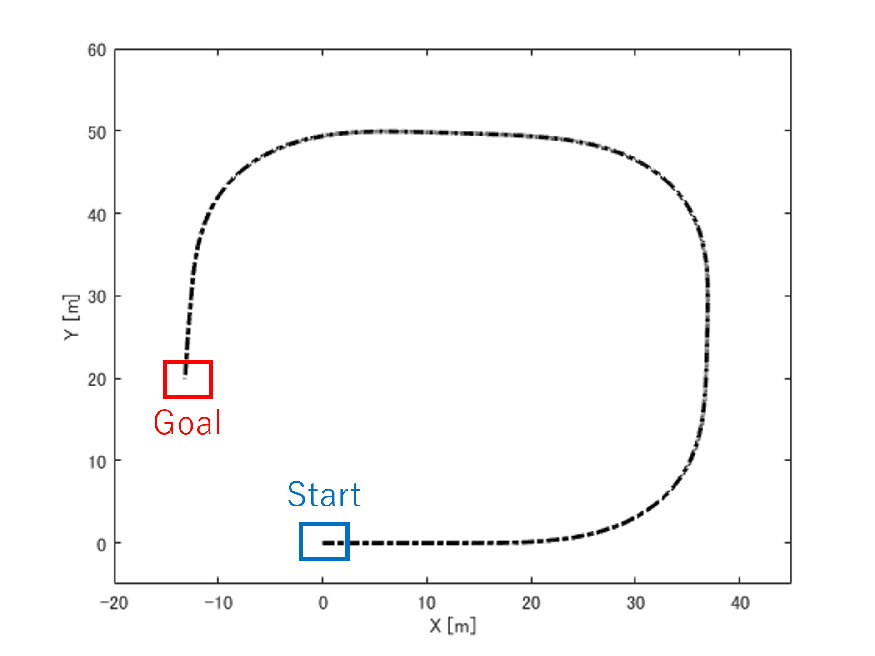}
\subcaption{Target path 1 (Square shaped path)}
\label{targetpath1}
\end{minipage}
\hfill
\begin{minipage}[t]{0.49\columnwidth}
\centering
\includegraphics[width=\columnwidth, height=4cm]{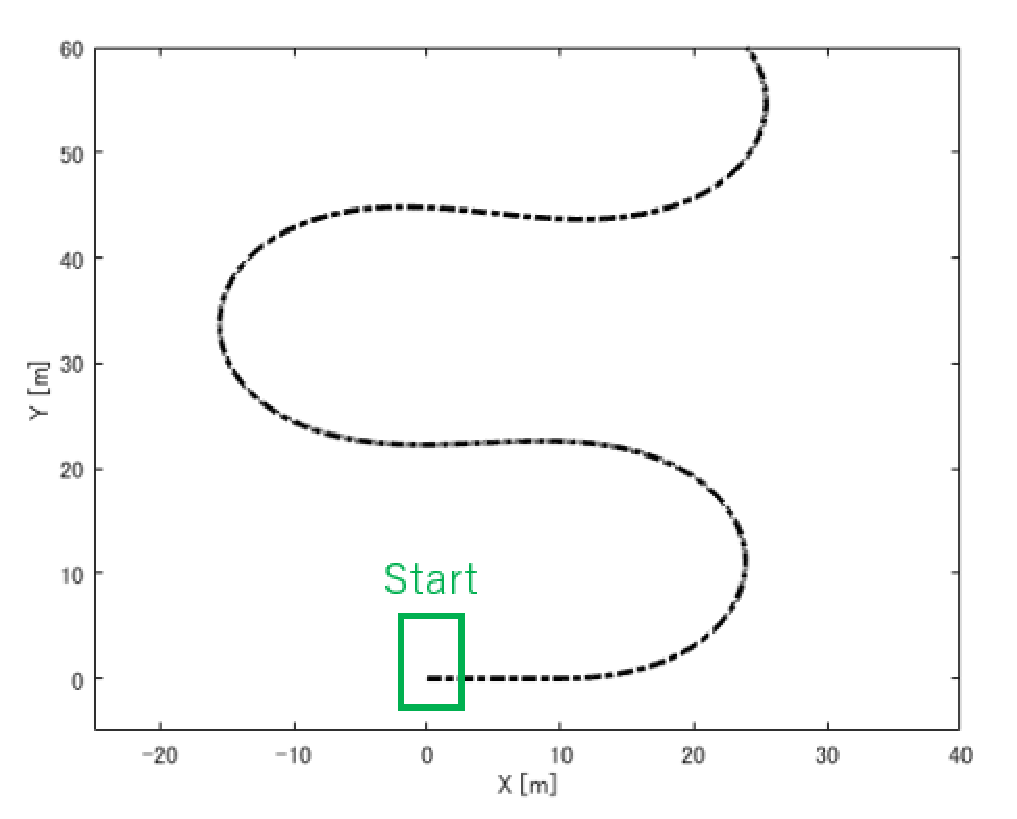}
\subcaption{Target path 2 (Meandering shaped path)}
\label{targetpath2}
\end{minipage}
\caption{Target paths used in the numerical simulations}
\label{fig:targetpaths}
\end{figure}

This section describes the two target paths used in the numerical simulations. First, the path defined by (\ref{4.1}) and illustrated in Fig.\,\ref{targetpath1} is used as target path $1$.
\begin{equation}
\kappa_{r}(s)=
\left\{ 
\begin{array}{ll}
0 & (0 \le s < 12) \\
0.037(1 - \cos(0.15s-1.8)) & (12 \le s \le 150)
\end{array} \right.
\label{4.1}
\end{equation}
In (\ref{4.1}), for $0$\,m to $12$\,m, the curvature of the target path, $\kappa_{r}$, is $0$, indicating a straight line segment. Beyond $12$\,m, the curvature of the target path gradually changes at each point, forming a curved trajectory. As a whole, the target path follows a square shaped pattern, as shown in Fig.\,\ref{targetpath1}.

Next, the path described by (\ref{4.2}) and illustrated in  Fig.\,\ref{targetpath2} is used as target path $2$.
\begin{equation}
\kappa_{r}(s)=
\left\{ 
\begin{array}{ll}
0 & (0 \le s < 5) \\
(3\sigma^{2}-2\sigma^{3})\kappa_{b}+(\sigma^{3}-\sigma^{2})L\kappa_{b}' & (5 \le s < 15) \\
0.1\sin(0.06s-0.2) & (15 \le s \le 207)
\end{array} \right.
\label{4.2}
\end{equation}
In (\ref{4.2}), the target path consists of a straight segment ($0 \le s < 5$\,m), a clothoid-like transition segment ($5 \le s < 15$\,m), and a sinusoidal segment ($15 \le s \le 207$\,m). The transition segment is a cubic Hermite interpolation, where $\sigma=(s-5)/L$ with $L=10$\,m, and $\kappa_{b}=0.1\sin(0.7)$ and $\kappa_{b}'=0.006\cos(0.7)$ are the curvature and its derivative of the sinusoidal segment at $s=15$\,m. This construction ensures that both $\kappa_{r}$ and $d\kappa_{r}/ds$ are continuous over the entire path, consistent with the smoothness assumption on the target path introduced in Section~\ref{sec2}. As a whole, the target path follows a meandering shaped pattern, as shown in Fig.\,\ref{targetpath2}.

The initial position and orientation are set as follows: $\xi(0)=0$, $\eta(0)=0$, and $\theta_o(0)=0$. The initial state of the plant is defined as $[\beta, \dot{\psi}, \delta, \theta, z]=[0, 0, 0, 0, 3]$ with $s_{r}(0)=0$, and the system follows the target path accordingly. The velocity of the plant is set to $v=3$\,m/s, and the wheelbase is $l=2.55$\,m. Also, the parameters of the vehicle are given by
\begin{equation}
\left[
\begin{array}{ccc}
a_{11} & a_{12} & a_{13}\\
a_{21} & a_{22} & a_{23}
\end{array}
\right]
=
\left[
\begin{array}{ccc}
-79.5 & 22.6 & 30.1\\
15.4 & -87.5 & 26.8
\end{array}
\right]
\end{equation}
These coefficients correspond to the following physical vehicle parameters: mass $m=1507$\,kg, front cornering power $K_{f}=45373$\,N/rad, rear cornering power $K_{r}=74406$\,N/rad, front axle distance $l_{f}=1.30$\,m, rear axle distance $l_{r}=1.25$\,m, and moment of inertia $I=2205$\,kg$\cdot$m$^{2}$, which are taken from \cite{A10}. Since $K_{r}>0$, the zero dynamics stability condition derived in Section~\ref{sec2-zerodynamics} is satisfied; the understeer index $K_{f}l_{f}/(K_{r}l_{r})\approx0.63<1$ is also confirmed for completeness.

Furthermore, the coefficients of the Hurwitz polynomial are set as  $\alpha_{1}=400$, $\alpha_{2}=500$, and $\alpha_{3}=240$. Following the MEC design principle of employing high-gain feedback in the frequency band requiring robustification \cite{A12}, the feedback gains $k_{1}$ to $k_{3}$ are chosen to be fifteen times these Hurwitz coefficients ($k_{i}=15\alpha_{i}$). This configuration ensures sufficiently fast error correction relative to the nominal convergence rate, effectively suppressing the impact of model errors and disturbances.

In this study, numerical simulations are conducted under two scenarios: one where the steering torque resistance coefficient of the model $C_{M}$ matches the steering torque resistance of the plant $C$, and another where they differ. For each scenario, we compare the simulation results between the conventional method and the proposed method. The conventional method, based on \cite{A10}, shares the identical path following controller structure with the proposed method and differs only in the absence of MEC compensation; this comparison therefore serves as an ablation evaluation that isolates the contribution of the MEC to the robustness against steering torque uncertainty.

\begin{remark}
We also investigated disturbance observer-based control (DOBC) \cite{A15} as an alternative benchmark. A classical DOBC implementation constructs the observer using an inverse of the nominal plant model; for the nonlinear path following system considered here, an exact inverse is not available, and only an approximate implementation is possible \cite{A15,A16}. In our preliminary implementation, although the disturbance observer rejected the matched disturbance at the steering input level, the tracking error grew without settling on curved path segments. We attribute this behavior to the curvature-dependent coupling in the path following kinematics, which propagates the residual error outside the input channel and is therefore difficult to compensate by input-level disturbance rejection alone. Since this observation is conditional on our specific implementation, a quantitative comparison of DOBC and the proposed method under a common controller setting is left for future work.
\end{remark}

\subsection{Simulation results}
\subsubsection{Simulation result for target path $1$}
\begin{figure}[b]
\centering
\includegraphics[width=8.0cm, height=7cm]{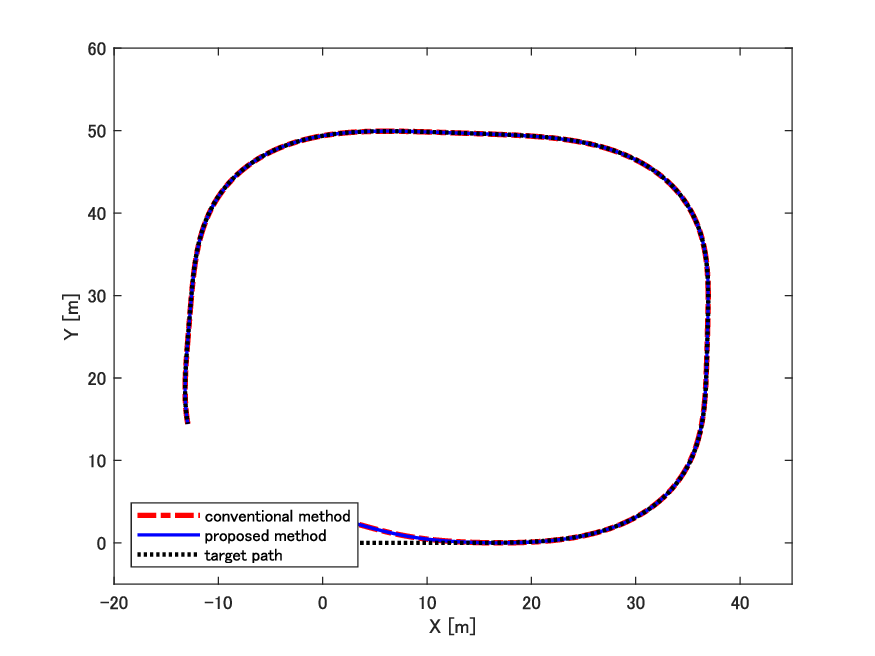}
\caption{Path following control with and without MEC in target path 1 ($C_{M}=200$, $C=200$)}
\label{Path following control with and without MEC(C=200)}
\end{figure}

First, we consider the case where the steering torque resistance coefficient of the model $C_{M}$ matches the steering torque resistance of the plant $C$. Since the zero dynamics are exponentially stable for the parameters used here, as established in Section~\ref{sec2-zerodynamics}, the stability of the closed-loop system is guaranteed when $C=C_{M}$. The path following results obtained using the conventional method and the proposed method are shown in Fig.\,\ref{Path following control with and without MEC(C=200)}.

In Fig.\,\ref{Path following control with and without MEC(C=200)}, the dashed line represents the target path, while the red solid line represents the trajectory of the conventional method and the blue solid line represents that of the proposed method. Both trajectories closely follow the target path, and the following errors of both methods remain at approximately $0.02$\,m after the initial transient. In fact, the red and blue lines completely overlap, so that only one line appears visible in the figure. These results are consistent with the theoretical expectation based on the stability of the zero dynamics.
\begin{figure}[t]
\centering
\includegraphics[width=8.0cm, height=7cm]{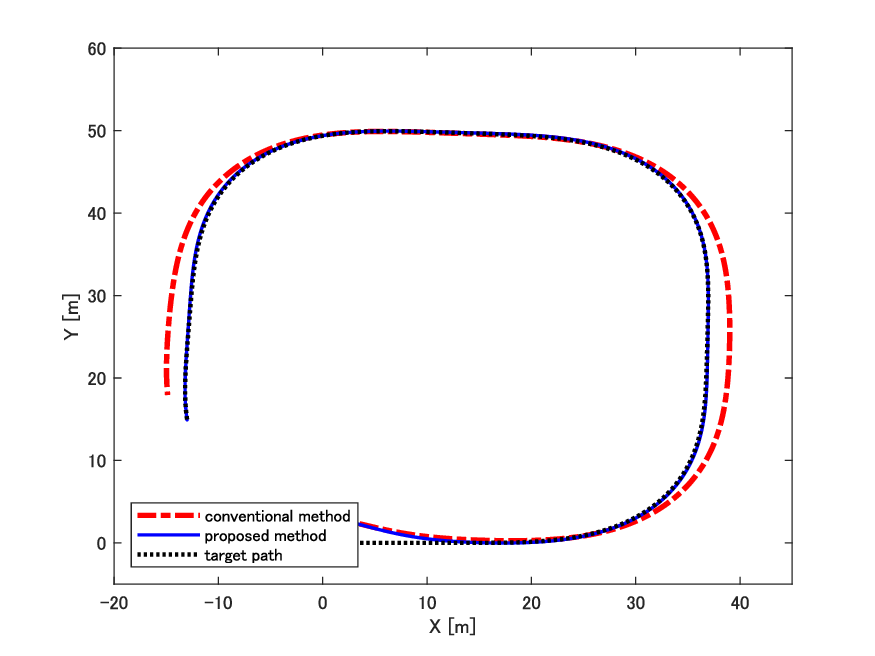}
\caption{Path following control with and without MEC in target path $1$ ($C_{M}=200$, $C=230$)}
\label{Path following control with and without MEC(C=230)}
\end{figure}
\begin{figure}[t]
\centering
\includegraphics[width=8.0cm]{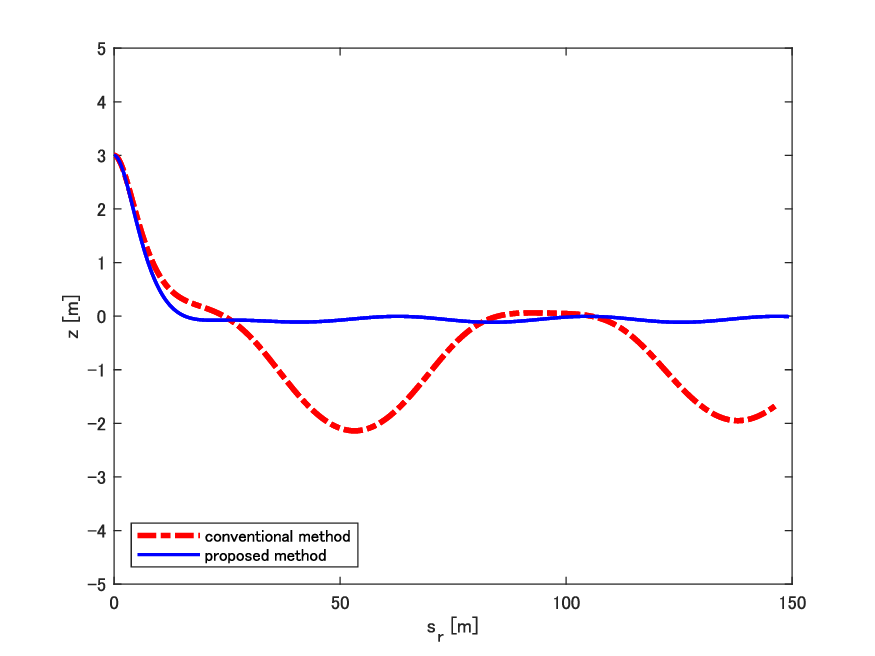}
\caption{Following error in target path $1$ ($C_{M}=200$, $C=230$)}
\label{Following error(C=230)}
\end{figure}

Next, we consider the case where the steering torque resistance coefficient of the model $C_{M}$ differs from the steering torque resistance of the plant $C$. As in the previous case, the simulation results obtained using the conventional method and the proposed method are shown in Fig.\,\ref{Path following control with and without MEC(C=230)}. The corresponding path following errors are depicted in Fig.\,\ref{Following error(C=230)}. In Fig.\,\ref{Path following control with and without MEC(C=230)}, the dashed line represents the target path, while the solid line represents the actual trajectory of the plant. In Fig.\,\ref{Following error(C=230)}, the horizontal axis denotes the path length, and the vertical axis represents the following error. The red line corresponds to the conventional method, while the blue line represents the proposed method.

\begin{figure}[t]
\centering
\includegraphics[width=8.0cm]{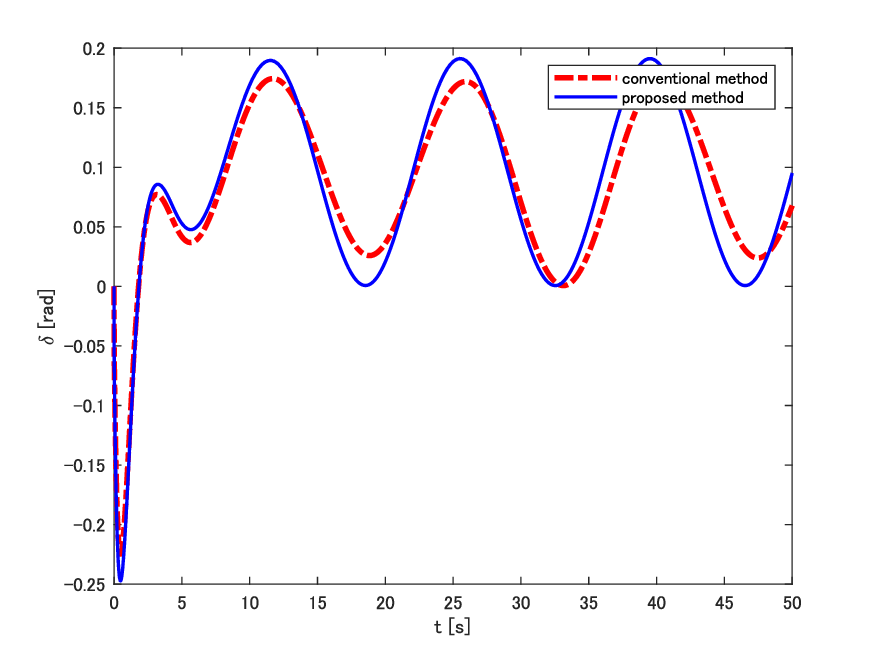}
\caption{Steering angle in target path $1$ ($C_{M}=200$, $C=230$)}
\label{fig:steering1}
\end{figure}
\begin{figure}[t]
\centering
\includegraphics[width=8.0cm]{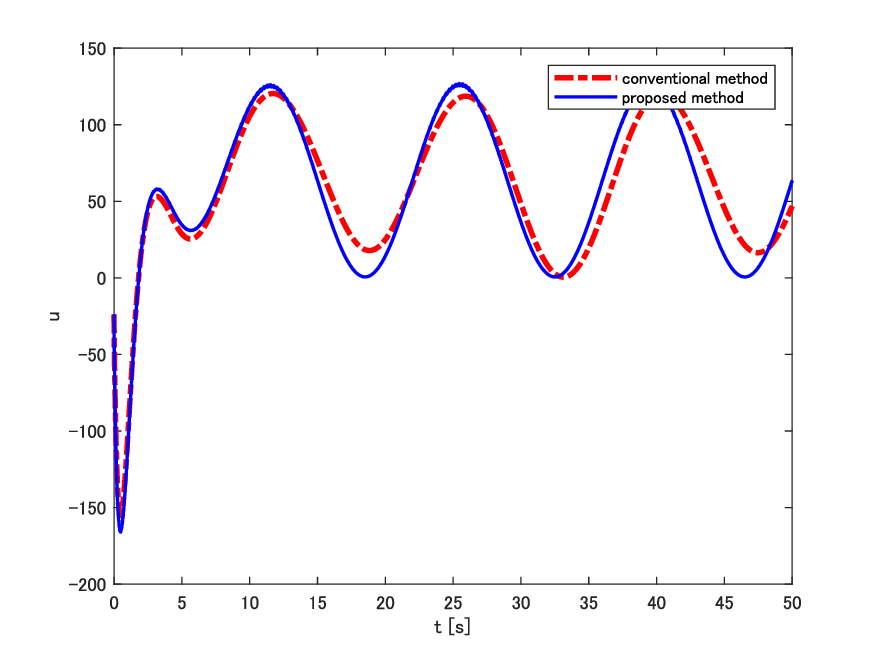}
\caption{Control input in target path $1$ ($C_{M}=200$, $C=230$)}
\label{fig:input1}
\end{figure}

In Fig.\,\ref{Path following control with and without MEC(C=230)}, when using the conventional method, the trajectory follows the target path closely, but the following error is relatively large. In contrast, the proposed method significantly reduces the following error. Specifically, in Fig.\,\ref{Following error(C=230)}, the red line shows a maximum following error of approximately $2.14$\,m, whereas the blue line shows a maximum following error of approximately $0.11$\,m. These results indicate that the application of MEC mitigates the effect of model mismatch.

To provide further insight into the control behavior, the time histories of the steering angle $\delta(t)$ and the control input $u(t)$ for the case $C_M=200$, $C=230$ are shown in Fig.\,\ref{fig:steering1} and Fig.\,\ref{fig:input1}, respectively. In the conventional method, the control input does not account for the mismatch in steering resistance, resulting in an insufficient steering angle to follow curved sections of the path. In contrast, the proposed method generates a larger control input through MEC compensation, effectively correcting the steering angle to reduce the tracking error. Note that the vehicle velocity is constant at $v=3$\,m/s throughout all simulations.

To quantitatively evaluate the robustness against parameter mismatch, the maximum following error is analyzed for various values of $C$. Here, the maximum following error refers to the peak value of $|z(t)|$ after the initial transient has settled, since the initial offset $z(0)=3$\,m is deliberately introduced to verify convergence and does not reflect steady-state tracking performance; specifically, the maximum is evaluated over the range $s_{r}\geq40$\,m. The relationship between the parameter ratio $C/C_{M}$ and the maximum following error is summarized in Table\,\ref{table:robustness1}.

From Table\,\ref{table:robustness1}, it can be observed that the following error is minimized when $C=C_{M}$ and increases as the parameter ratio $C/C_{M}$ deviates from unity. This trend is observed for both methods; however, the increase in error is more pronounced in the conventional method. In contrast, the proposed method suppresses the increase in following error under parameter mismatch. Based on these results, the proposed method maintains relatively good tracking performance within the range $0.75\leq C/C_{M}\leq1.25$, while the performance gradually degrades beyond this range. Furthermore, when the parameter mismatch becomes large, the proposed method also exhibits gradual performance degradation, although its tracking error remained bounded over the entire tested range. Although a rigorous stability analysis is not conducted in this study, these numerical observations provide practical insight into the allowable range of parameter mismatch for the proposed method.

\begin{table}[t]
\centering
\caption{Maximum following error under parameter mismatch in target path $1$}
\label{table:robustness1}
\scalebox{1.2}{
\begin{tabular}{|c|c|c|c|}
\hline
$C$ & $C/C_{M}$ & Conventional [m] & Proposed [m] \\
\hline
100 & 0.5 & 17.1 & 0.47 \\
150 & 0.75 & 4.14 & 0.25 \\
200 & 1.0 & 0.02 & 0.02 \\
250 & 1.25 & 3.47 & 0.20 \\
300 & 1.5 & 6.51 & 0.43 \\
400 & 2.0 & 11.5 & 0.87 \\
\hline
\end{tabular}
}
\end{table}

\subsubsection{Simulation result for target path $2$}
\begin{figure}[t]
\centering
\includegraphics[width=8.0cm, height=7cm]{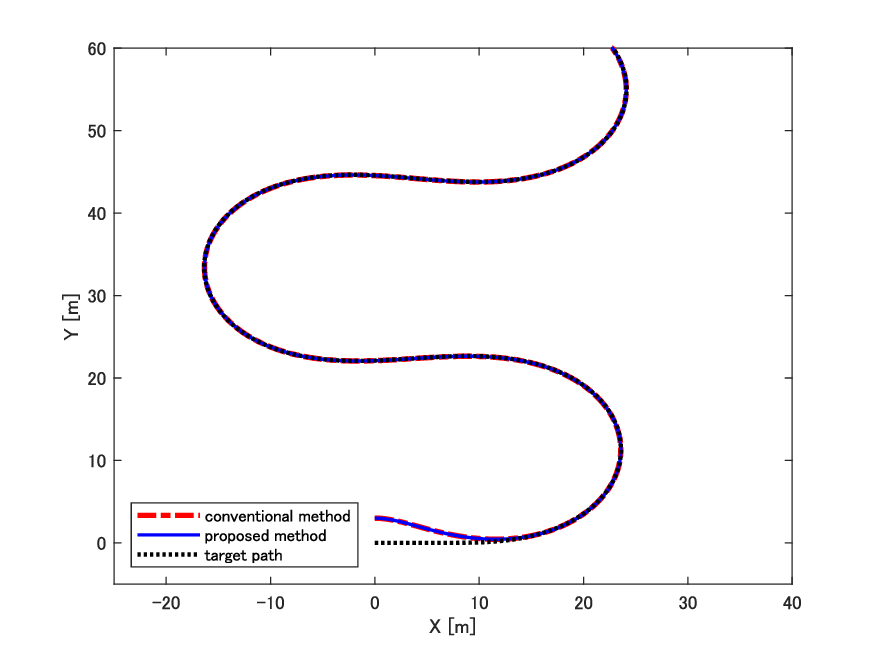}
\caption{Path following control with and without MEC in target path $2$ ($C_{M}=200$, $C=200$)}
\label{Path following control with and without MEC2(C=200)}
\end{figure}

First, we consider the case where the steering torque resistance coefficient of the model $C_{M}$ matches the steering torque resistance of the plant $C$. As in the previous case, the stability of the closed-loop system is guaranteed when $C=C_{M}$ owing to the stability of the zero dynamics (Section~\ref{sec2-zerodynamics}). The path following results obtained using the conventional method and the proposed method are shown in Fig.\,\ref{Path following control with and without MEC2(C=200)}.

In Fig.\,\ref{Path following control with and without MEC2(C=200)}, both trajectories closely follow the target path, with following errors remaining at approximately $0.03$\,m for both methods after the initial transient. The results of the two methods completely overlap, consistent with those obtained for target path~$1$.

\begin{figure}[t]
\centering
\includegraphics[width=8.0cm, height=7cm]{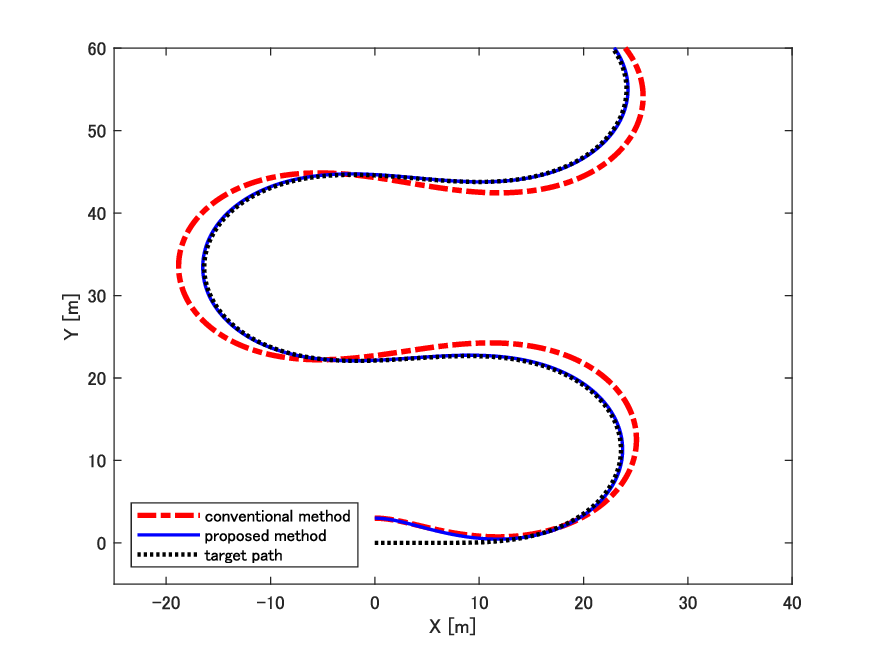}
\caption{Path following control with and without MEC in target path $2$ ($C_{M}=200$, $C=230$)}
\label{Path following control with and without MEC2(C=230)}
\end{figure}

\begin{figure}[t]
\centering
\includegraphics[width=8.0cm]{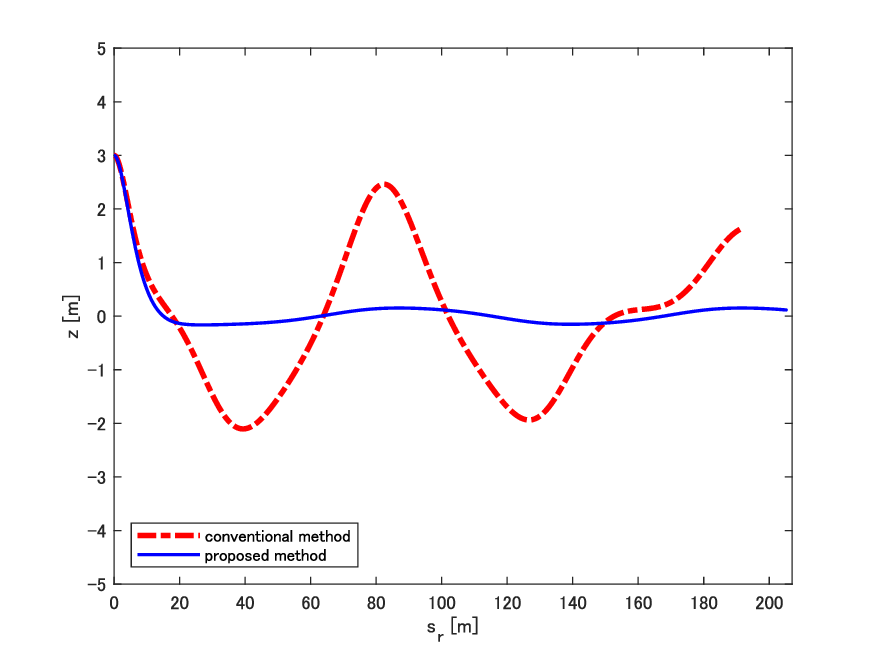}
\caption{Following error in target path $2$ ($C_{M}=200$, $C=230$)}
\label{Following error2(C=230)}
\end{figure}

Next, we consider the case where $C_{M}$ differs from $C$. The simulation results obtained using the conventional method and the proposed method are shown in Fig.\,\ref{Path following control with and without MEC2(C=230)}, and the corresponding path following errors are depicted in Fig.\,\ref{Following error2(C=230)}. In Fig.\,\ref{Path following control with and without MEC2(C=230)}, the conventional method exhibits relatively large deviations from the target path. In contrast, the proposed method significantly reduces the following error. Specifically, the maximum following error is approximately $2.46$\,m for the conventional method, whereas it is reduced to approximately $0.15$\,m with the proposed method. These results indicate that the effectiveness of MEC is maintained across different target paths.

As in target path~$1$, the proposed method generates appropriate compensation through MEC, adjusting the steering torque to maintain accurate path following despite the parameter mismatch.

In addition, the relationship between the parameter ratio $C/C_{M}$ and  the maximum following error is illustrated in Table\,\ref{table:robustness2}. 


From Table\,\ref{table:robustness2}, a similar trend can be observed: the following error is minimized when $C=C_{M}$ and increases as the parameter mismatch becomes larger. This tendency is common to both methods; however, the increase in error is more pronounced in the conventional method. In contrast, the proposed method suppresses the increase in following error under parameter mismatch, showing results consistent with those obtained for target path $1$. Based on these results, the proposed method maintains relatively good tracking performance within the range $0.75\leq C/C_{M}\leq1.25$, while the performance gradually degrades beyond this range. Furthermore, even in the extreme case $C/C_{M}=0.5$, where the conventional method diverges, the proposed method maintains a bounded tracking error of approximately $0.65$\,m, although the error grows as the mismatch increases. Although a rigorous stability analysis is not conducted in this study, these numerical observations provide practical insight into the allowable range of parameter mismatch and confirm the consistency of the results across different target paths.

\begin{table}[b]
\centering
\caption{Maximum following error under parameter mismatch in target path $2$}
\label{table:robustness2}
\scalebox{1.2}{
\begin{tabular}{|c|c|c|c|}
\hline
$C$ & $C/C_{M}$ & Conventional [m] & Proposed [m] \\
\hline
100 & 0.5 & Diverged & 0.65 \\
150 & 0.75 & 5.00 & 0.33 \\
200 & 1.0 & 0.03 & 0.03 \\
250 & 1.25 & 3.98 & 0.27 \\
300 & 1.5 & 7.40 & 0.56 \\
400 & 2.0 & 13.2 & 1.14 \\
\hline
\end{tabular}
}
\end{table}

From the results above, it can be observed that when the steering torque resistance coefficient of the model $C_{M}$ matches that of the plant $C$, both methods achieve accurate path following. When $C_{M}$ and $C$ differ, the conventional method shows a noticeable degradation in tracking performance, whereas the proposed method reduces the following error by compensating for the model error caused by parameter mismatch. These results indicate that the proposed method improves robustness against parameter mismatch and maintains stable tracking performance over a range of conditions.

\section{Conclusion}
This paper has proposed a robust path-following control system for vehicles by incorporating steering dynamics and compensating for model errors using a Model Error Compensator (MEC). The control framework was developed based on a vehicle model that includes slip angle, yaw rate, and steering angle as state variables, with the vehicle velocity assumed to be constant. The zero dynamics associated with the path following formulation were formally derived, and it was shown that their stability is guaranteed whenever the rear cornering power is positive ($K_{r}>0$), a simple condition that holds for both understeer and oversteer configurations, providing the theoretical foundation for the proposed control design. Considering practical challenges such as variations in steering resistance due to environmental conditions, the proposed method aims to enhance robustness in path-following performance.

Numerical simulations were conducted to validate the effectiveness of the proposed method under conditions where the assumed model parameters matched or differed from the actual values. The results demonstrated that, while conventional control laws perform adequately when model parameters are accurate, the MEC-based approach significantly improves tracking performance when discrepancies exist. The application of MEC effectively suppressed the influence of model uncertainties, leading to reduced path-following errors and improved robustness. Furthermore, simulation results (Table\,\ref{table:robustness1} and Table\,\ref{table:robustness2}) confirmed that practical tracking performance is maintained within a parameter variation range of $0.75 \leq C/C_{M} \leq 1.25$.

However, this study is limited to simulation-based validation, and a rigorous stability analysis of the closed-loop system has not been conducted. In addition, the steering torque model $Cv\delta$ used in this study is a simplified representation, and extension to a more realistic nonlinear model is required. Moreover, while acceptable tracking performance was observed within the aforementioned parameter range, theoretical guarantees of stability outside this range remain an open issue. Furthermore, since the vehicle velocity is assumed to be constant, the behavior under acceleration and deceleration conditions has not been evaluated.

In future work, the most immediate priority is a rigorous stability analysis of the closed-loop system under parameter mismatch, building on the zero dynamics characterization in Section~\ref{sec2-zerodynamics}, including a formal characterization of the allowable range of mismatch. A promising direction for this analysis is contraction theory \cite{A25}, which characterizes the exponential convergence of neighboring trajectories of a system: since MEC operates by feeding back the discrepancy between the actual plant and the nominal model running in parallel, the boundedness of the tracking error can be studied by analyzing the contraction property of the error dynamics between these two trajectories under the high-gain compensation. Building on this theoretical foundation, the framework will be extended to account for velocity variations. In addition, a quantitative comparison with other classes of robust control approaches, such as DOBC and ADRC, implemented under a common controller setting, remains an important direction for future work. Furthermore, toward path following on uneven road surfaces, extending the present planar formulation to three-dimensional paths, building on recent path following formulations such as \cite{A24}, is also envisioned.

Experimental validation using a physical vehicle platform is also planned as a critical next step. Specifically, the proposed method will be implemented on a low-speed autonomous vehicle equipped with a steering torque sensor and encoder-based state measurement. The experiments will be conducted under multiple road surface conditions (e.g., asphalt, gravel, and soil) to evaluate the practical effectiveness of MEC compensation against real-world variations in steering torque resistance. A particular focus will be placed on agricultural vehicles operating on unstructured terrain, where vehicle-specific parameter identification and the influence of varying soil conditions must be addressed. While the present simulation study has demonstrated the conceptual validity and robustness characteristics of the proposed method under controlled conditions, experimental verification is essential to confirm its applicability in practical scenarios.

\end{document}